\documentclass[11pt]{article}

\usepackage[final]{acl}

\usepackage{times}
\usepackage{latexsym}

\usepackage[T1]{fontenc}

\usepackage[utf8]{inputenc}

\usepackage{microtype}

\usepackage{inconsolata}

\usepackage{graphicx}

\usepackage{booktabs}

\title{Evaluating Pre-trained Speech Encoders for Spontaneous Speech Detection and Out of Domain Synthetic Speech Generalisation in Indic Languages}

\author{
  \textbf{Varun Rai\textsuperscript{1}},
  \textbf{Pavan Kumar J\textsuperscript{2}},
  \textbf{Sujith Pulikodan\textsuperscript{2}},
  \textbf{Nihar Desai\textsuperscript{2}} \\
  \textsuperscript{1}Indian Institute of Technology Guwahati,Assam, India \\
  \textsuperscript{2}AI \& Robotics Technology Park (ARTPARK), I-Hub @ IISc, Bangalore, India\\ 
  \small{
    \textbf{} \href{mailto:varun.rai@op.iitg.ac.in }{varun.rai@op.iitg.ac.in }
  }
}

\begin{document}
\maketitle
\begin{abstract}
Transformer-based models have shown strong accuracy in distinguishing spontaneous from scripted speech and natural from synthetic speech, but these results are established on a narrow set of well-resourced language benchmarks and have not been extended across Indic languages, nor has embedding geometry been used to explain encoder behaviour or deepfake generalisation failure. We address these gaps by evaluating five frozen transformer encoders, AST, Vaani-FastConformer, Wav2vec2, Whisper and BEATs, across 22 Indic languages, and by conducting a multi-system TTS generalisation experiment across four TTS models. Beyond accuracy, we present language isolation probing and centroid proximity analysis. Probing reveals an encoder-dependent trade-off between language-discriminability and spontaneity detection. Centroid analysis shows that out-of-domain generalisation is predicted by a training system's proximity to unseen TTS embeddings, not its distance from natural speech, a finding with direct implications for training data selection in real-world deepfake detectors.
\end{abstract}

\section{Introduction}

The rapid improvement of neural text-to-speech (TTS) and voice conversion (VC) systems has transformed synthetic speech into a near-indistinguishable representation of human voice, opening new vectors for impersonation, disinformation, and fraud. Distinguishing natural from synthetic speech has become a substantive research problem, and distinguishing spontaneous from scripted natural speech is equally important for downstream applications ranging from conversational AI to emotion recognition.

Research on both tasks has made significant progress, but that progress is linguistically concentrated. The ASVspoof \citep{asvspoof2019,asvspoof2021,asvspoof2023} challenge series, which has anchored audio deepfake detection since 2015, operates in English-centric settings with evaluation data dominated by high-resource languages. The literature on spontaneous versus scripted speech classification --- anchored by prosody-oriented studies \citep{batliner1995} and, more recently, transformer-based classifiers such as \citet{elisha2024} --- has similarly been benchmarked on European languages, podcast corpora, or acted datasets such as IEMOCAP \citep{busso2008}, none of which reflects the typological diversity or low-resource conditions characteristic of Indian languages.

The detection field has itself undergone a significant architectural shift. Handcrafted front-ends such as CQCC \citep{todisco2017cqcc} and LFCC \citep{wu2015lfcc} have given way to large pre-trained speech foundation models --- Wav2vec2 \citep{baevski2020wav2vec2}, Whisper \citep{radford2022whisper}, WavLM \citep{chen2022wavlm} --- whose frozen representations substantially outperform task-specific architectures such as AASIST \citep{jung2022} and RawNet2 \citep{tak2021} on out-of-distribution synthetic speech \citep{li2024}. Yet generalisation remains conditioned on training diversity: detectors trained on a narrow synthetic distribution generalise poorly even with powerful encoder backbones, and prior work benchmarking AASIST and RawNet2 on Indic synthetic speech \citep{sharma2025} has observed near-chance performance.

We address these gaps with a unified study across both tasks over 22 Indic languages. Our contributions are: (1) the first systematic evaluation of five frozen transformer encoders on spontaneous versus read classification across all 22 scheduled Indian languages; (2) a language isolation probing analysis revealing an encoder-dependent trade-off between language-discriminability and spontaneity detection; (3) a multi-system TTS generalisation experiment showing out-of-domain synthetic recall rising from 7\% to 51\% as the training pool expands from one to four Indic TTS systems; and (4) an embedding centroid analysis showing that generalisation is predicted by a training system's proximity to unseen TTS embeddings rather than its distance from natural speech.

\section{Spontaneous and Scripted Speech Classification}

\subsection{Related Work}

The acoustic differences between spontaneous and read speech have been studied for decades, with prosody consistently identified as a primary discriminating dimension. Early work by \citet{batliner1995} showed pitch contours, articulation rate, and pause distribution carry signal, though no single feature dominates. \citet{rouas2003} found these prosodic differences large enough to affect downstream task performance, and \citet{nakamura2008} showed ASR models trained on read speech degrade substantially on spontaneous input. \citet{christodoulides2020} and \citet{ryant2016} each built multidimensional characterisations of speaking styles from automatically-extracted prosodic features. \citet{christodoulides2020} found spontaneous conversational speech shows markedly higher individual variability than more homogeneous styles, making it the hardest to characterise; \citet{ryant2016} found pitch range, unlike segment-duration measures, is unreliable for distinguishing spontaneous from read speech, since speakers vary widely in pitch range regardless of style. This asymmetry --- read speech is easier to classify than spontaneous --- appears consistently across the literature.

A direct precursor is \citet{mangalam2018}, who approached the spontaneous versus scripted distinction using the IEMOCAP corpus, achieving approximately 80\% five-fold cross-validation accuracy on the binary task with SVM-based classifiers, and showing that incorporating spontaneity detection as an auxiliary task improved emotion recognition accuracy by about 3 percentage points. The most directly relevant prior work is \citet{elisha2024}, who conducted the first systematic comparison of modern audio transformers---including Whisper and YAMNet---against handcrafted features on a large multilingual podcast dataset covering eleven language groups. They found transformer-based models consistently outperform handcrafted baselines, but their work does not include embedding analysis and centres on high resource languages.

\subsection{Dataset}

The primary source of natural Indic speech is IndicVoices \citep{javed2024}, a large-scale multilingual corpus covering 22 scheduled Indian languages collected from speakers across 208 districts of India. Each recording carries a scenario tag Read, Extempore or Conversation. This scenario column serves as clean binary ground truth for the spontaneous versus read task. Subset from the train set is used for training, and a subset from the valid set is used for testing.
 
IEMOCAP. The Interactive Emotional Dyadic Motion Capture (IEMOCAP) database \citep{busso2008} consists of approximately 12 hours of audiovisual recordings from ten actors performing across five dyadic sessions. Each session contains two interaction types: scripted scenes in which actors deliver pre-written emotional dialogues, and improvised scenes in which actors respond spontaneously to emotionally charged hypothetical scenarios. The interaction type is encoded directly in the filename of each recording: files containing \_script belong to the scripted (read) condition and those containing \_impro belong to the improvised (spontaneous) condition. This filename convention is the sole source of ground truth labels for all IEMOCAP utterances in our experiments. The five sessions are numbered 1 through 5; sessions 1 to 4 contribute to training and validation, while session 5 is held out exclusively for testing, providing full speaker independence since each session involves a unique pair of actors.

\subsection{Classifier Architecture}

A compact three-block fully connected DNN is trained on frozen encoder embeddings. The 768-D variant (Whisper-small \citep{radford2022whisper}, AST \citep{gong2021ast}, BEATs \citep{chen2023beats}) uses dimensions 768 $\rightarrow$ 192 $\rightarrow$ 64 $\rightarrow$ 1, yielding 281,330 trainable parameters. The 1024-D variant (Wav2Vec2-large \citep{baevski2020wav2vec2}, Vaani-multilingual FastConformer \citep{pulikodan2026vaani}) uses dimensions 1024 $\rightarrow$ 128 $\rightarrow$ 64 $\rightarrow$ 1, yielding 323,030 parameters. This adjustment keeps classifier capacity broadly comparable across variants, ensuring that performance differences across encoders reflect upstream representation quality rather than downstream classifier capacity.

\subsection{Results: Spontaneous and read speech classification}

Figure 1 Reports spontaneous versus read speech accuracy across the four models and all 23 evaluation conditions, including 22 Indic language settings and IEMOCAP as the English speech reference. One aggregate row is included: AverageIndic reports the mean accuracy over all 22 Indic languages. It highlights that Whisper and Vaani outperform others across Indic languages as well as on English, with consistently higher accuracies.

\begin{figure}
    \centering
    \includegraphics[width=1.0\linewidth]{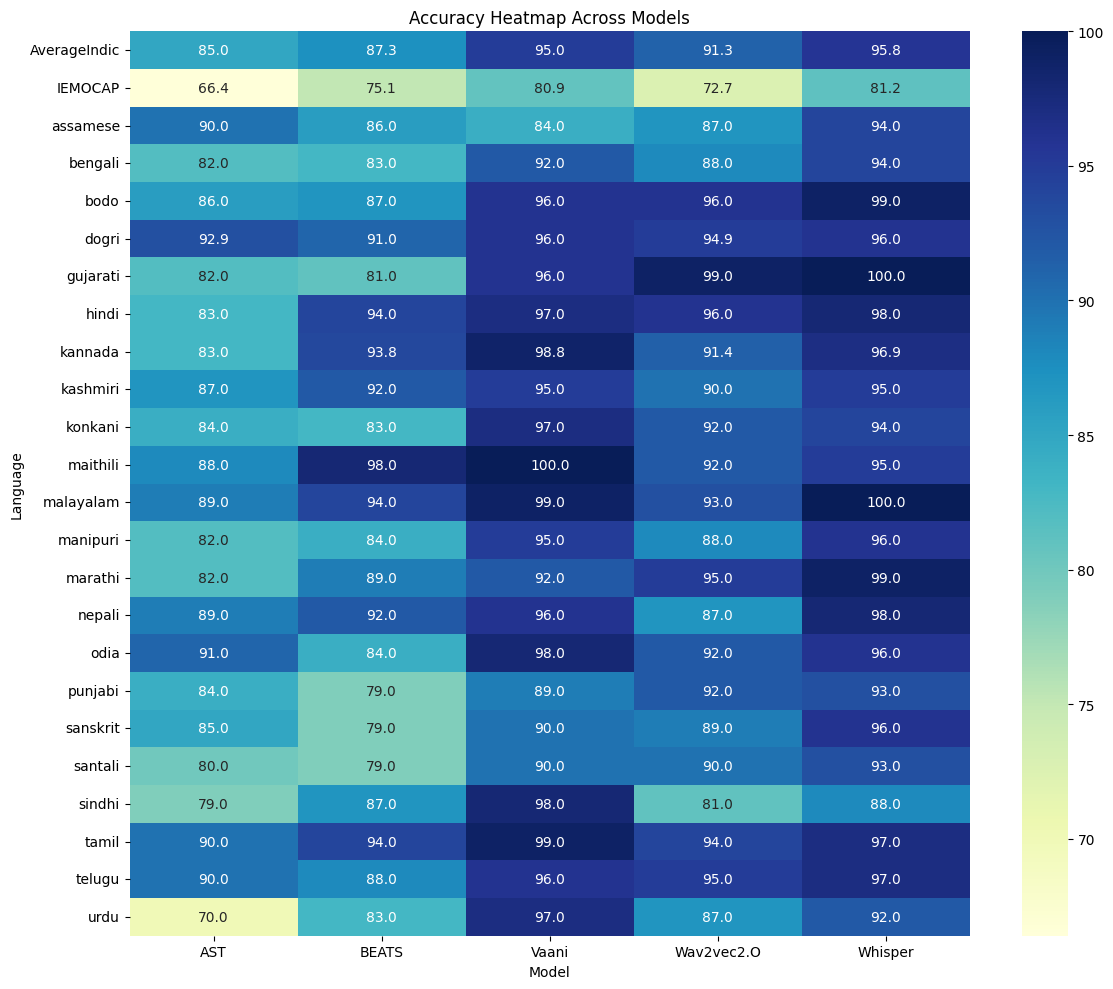}
    \caption{Accuracy heatmap}
    \label{fig:accuracy_heatmap}
\end{figure}

\subsubsection{Language wise invariance analysis}

The analysis asks a complementary question about the internal structure of those embeddings: to what extent do they retain explicit language identity information, and does that language-discriminability trade off against the model's ability to classify spontaneous speech?

For each encoder's feature store, all embedding matrices and their language labels are loaded and pooled. A multinomial logistic regression probe is trained to predict the language of origin from the embedding vector, using 70\% of a held-out 20\% validation partition for training the probe and 30\% for evaluation. The per-class recall of this probe --- referred to as the language isolation score --- measures how reliably the representations of utterances in language L can be identified as belonging to L
A high isolation score indicates that language-specific structure remains strongly encoded in the representation. 
The Pearson and Spearman correlations between the language isolation score and downstream spontaneity classification accuracy are then computed. Figure 2 shows the resulting scatter plots.

\begin{figure}
    \centering
    \includegraphics[width=1.0\linewidth]{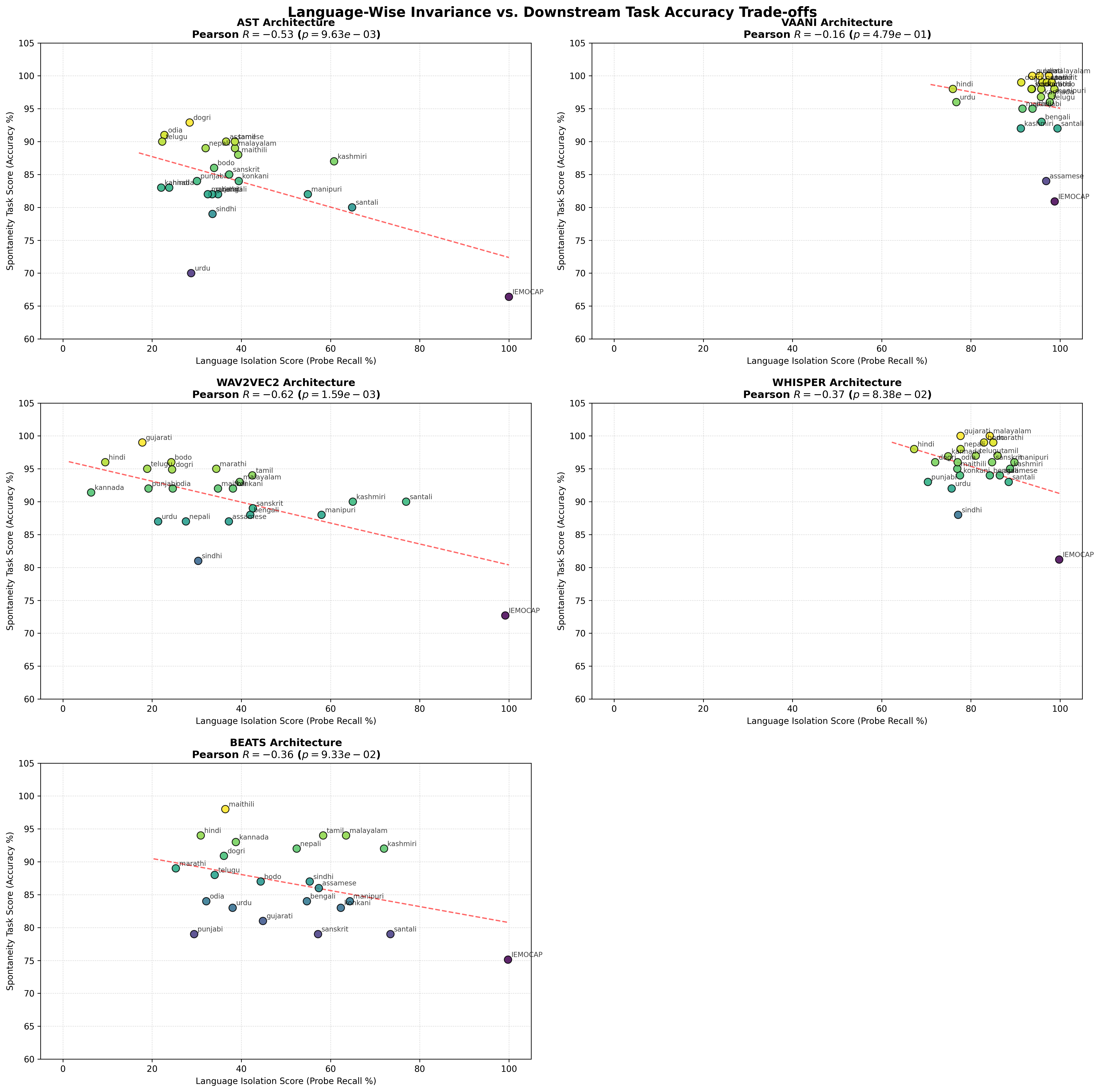}
    \caption{Language invariance plot}
    \label{fig:language_invariance}
\end{figure}

The pattern across models is divergent, and this divergence is the most interpretively rich finding from the post-training analyses.

For Wav2vec2 \citep{baevski2020wav2vec2}, both Pearson ($R=-0.62$, $p=1.50\times 10^{-3}$) and Spearman ($R=-0.48$, $p=2.12\times 10^{-2}$) correlations are significant and moderately strong. This means that for Wav2vec2, languages whose utterances are more easily identified by the probe --- i.e., whose embeddings are more language-discriminative - tend to achieve lower downstream spontaneity accuracy.

For AST \citep{gong2021ast}, Pearson is significant ($R=-0.52$, $p=0.01$) but Spearman is not ($R=-0.23$, $p=0.28$), suggesting the correlation is driven by structurally extreme outliers, most prominently IEMOCAP at near-100\% isolation. For BEATs \citep{chen2023beats}, neither correlation reaches significance (Pearson $R=-0.36$, $p=9.33\times 10^{-2}$), with languages spanning moderate isolation scores (20--65\%) and variable accuracy (79--97\%), indicating a noisy intermediate pattern distinct from Wav2vec2's clean trade-off.

For Whisper \citep{radford2022whisper} (Pearson $R=-0.37$, $p=0.083$) and Vaani \citep{pulikodan2026vaani} (Pearson $R=-0.15$, $p=0.485$), no significant correlation exists. Both encoders maintain consistently high accuracy regardless of how language-discriminative their representations are --- a flat high-accuracy plateau indicating that spontaneity detection in these encoders is decoupled from language-specific structure in the embedding space.

\section{Natural and Synthetic Speech Classification}

\subsection{Related Work}
Architectural progress in audio deepfake detection has been led by end-to-end models such as RawNet2 \citep{tak2021} and AASIST \citep{jung2022}, which set strong baselines on the ASVspoof \citep{asvspoof2019,asvspoof2021,asvspoof2023} benchmarks. However, their generalisation to unseen languages is severely limited: \citet{sharma2025} benchmarked both models on IndicSynth without domain adaptation and observed EERs exceeding 50\% across most Indic languages, compared to sub-1\% EER on ASVspoof 2019, exposing a critical cross-lingual generalisation gap. \citet{li2024} demonstrated through systematic analysis across multiple datasets that speech foundation models --- Whisper \citep{radford2022whisper} and Wav2Vec2 \citep{baevski2020wav2vec2} among them --- consistently outperform AASIST and RawNet2 on out-of-distribution synthetic speech, with Wav2Vec2BERT \citep{barrault2023seamless}  surpassing AASIST by over 20\% on unseen TTS systems. We adopt this finding as our rationale: rather than fine-tuning a task-specific detector, we extract frozen Whisper encoder embeddings via mean pooling and train a lightweight neural classifier.

\subsection{Dataset}
The training corpus comprises natural speech samples from IndicVoices and synthetic speech generated by four primary TTS systems: Indic F5 (F5) \citep{indicf52025}, Indic VITS (vitsrasa13) \citep{vitsrasa2024}, OmniVoice (Omni) \citep{zhu2026omnivoice}, and Meta M4 (M4) \citep{barrault2023seamless}. Each system synthesized audio using the same text prompts and corresponding natural recordings. For up to nine Indic languages, we generated 1,000 synthetic utterances per language per model. Of these, 200 utterances per language per model were reserved as the held-out test set, while the remaining 800 per language per model were used in varying combinations for training.

To evaluate generalization beyond the training distribution, we additionally incorporated two external TTS systems --- freevc24 \citep{li2023freevc} and xttsv2 \citep{casanova2024xtts} --- sourced from the IndicSynth \citep{sharma2025} dataset. These systems were excluded from classifier training and used solely for out-of-distribution (OOD) evaluation and embedding geometry analysis. For each language and model, 1,000 synthetic utterances were collected, forming a fully unseen OOD test partition.

\subsection{Classifier Architecture}

We train a compact three-block fully connected DNN on frozen Whisper-small embeddings \citep{radford2022whisper}. The network applies successive transformations of 768$\rightarrow$192$\rightarrow$64$\rightarrow$1, resulting in 281,330 trainable parameters.

\subsection{Classification Results and Embedding Space analysis}

To understand the structural properties underlying classifier behaviour, we computed centroid-level statistics on the 768-dimensional Whisper-small encoder embedding space. Table 2 reports cosine distance and Euclidean distance of each system's mean vector from all other system's centroid, Table 1 presents the recall and Figure 3 presents the t-SNE plot which shows clear clustering of synthetic speech by TTS system, with natural samples forming a distinct region.

\begin{table*}[t]
\centering
\small
\resizebox{\textwidth}{!}{%
\begin{tabular}{lcccccc}
\toprule
\hline
\textbf{Training Set} & \textbf{IndicSynth} & \textbf{Natural} & \textbf{F5} & \textbf{Omni} & \textbf{Indic VITS} & \textbf{M4} \\
\midrule
\hline
800 F5 & 0.0759 & 0.9939 & 0.9939 & 0.8127 & 0.0350 & 0.0567 \\
800 Omni & 0.2861 & 0.9939 & 0.6676 & 0.9972 & 0.0833 & 0.2050 \\
800 Indic VITS & 0.1065 & 1.0000 & 0.0050 & 0.0067 & 1.0000 & 0.5833 \\
800 M4 & 0.0889 & 0.9978 & 0.0028 & 0.0250 & 0.5722 & 1.0000 \\
400 Indic VITS + 400 M4 & 0.2251 & 0.9983 & 0.0072 & 0.0189 & 1.0000 & 1.0000 \\
400 F5 + 400 Indic VITS & 0.2560 & 0.9956 & 0.9878 & 0.6676 & 1.0000 & 0.7100 \\
400 Omni + 400 F5 & 0.3036 & 0.9889 & 0.9917 & 0.9972 & 0.1300 & 0.3000 \\
400 Omni + 400 M4 & 0.3267 & 0.9917 & 0.6954 & 0.9989 & 0.2906 & 1.0000 \\
267 F5 + 267 Omni + 267 Indic VITS & 0.4313 & 0.9944 & 0.9822 & 0.9933 & 1.0000 & 0.6383 \\
267 Omni + 267 Indic VITS + 267 M4 & 0.4637 & 0.9944 & 0.6481 & 0.9933 & 1.0000 & 1.0000 \\
\textbf{200 M4 + 200 Indic VITS + 200 Omni + 200 F5} & 
\textbf{0.5113} & 
\textbf{0.9911} & 
\textbf{0.9794} & 
\textbf{0.9972} & 
\textbf{0.9994} & 
\textbf{1.0000} \\
400 Omni + 400 Indic VITS & 0.4911 & 0.9939 & 0.7160 & 0.9939 & 1.0000 & 0.8417 \\
\bottomrule
\hline
\end{tabular}}
\caption{Summarises performance of all tested training configurations across 6 evaluation sets. 400 Omni + 400 Indic VITS is one of the training configurations, which means 400 synthetic audio files from both models were used to train the classifier, similar naming convention is followed by all other training configurations. All training sets include 800 natural audio samples.}
\label{tab:training_results}
\end{table*}

\subsection{Key Findings}

\textbf{1.} Single-system training yields poor OOD generalisation on unseen TTS systems, confirming the well-known brittleness of deepfake detectors trained on a narrow synthetic distribution. A performance ceiling of ~51\% OOD synthetic recall persists across all tested configurations, pointing to a distributional gap between training and evaluation systems that cannot be overcome by composition alone within the current system pool.

\textbf{2. }Training diversity is the primary lever for generalisation: expanding from one to four synthetic TTS systems in the training pool improves OOD synthetic recall from ~7\% to ~51\%, Greatest gains arise from including at least one training system whose embedding centroid lies proximate to the OOD evaluation region.
Omni Euclidean distances from xttsv2 and freevc24 embeddings are 1.26 and 1.91, respectively, which helped to improve the accuracy and on the other hand, F5 has closest embeddings to Natural embeddings (1.121), but its performance is worse on OOD synthetic recall. This analysis shows that out-of-domain generalisation is predicted by a training system's proximity to unseen TTS embeddings, not its distance from natural speech, a finding with direct implications for training data selection in real-world deepfake detectors.

\begin{figure}
    \centering
    \includegraphics[width=1.0\linewidth]{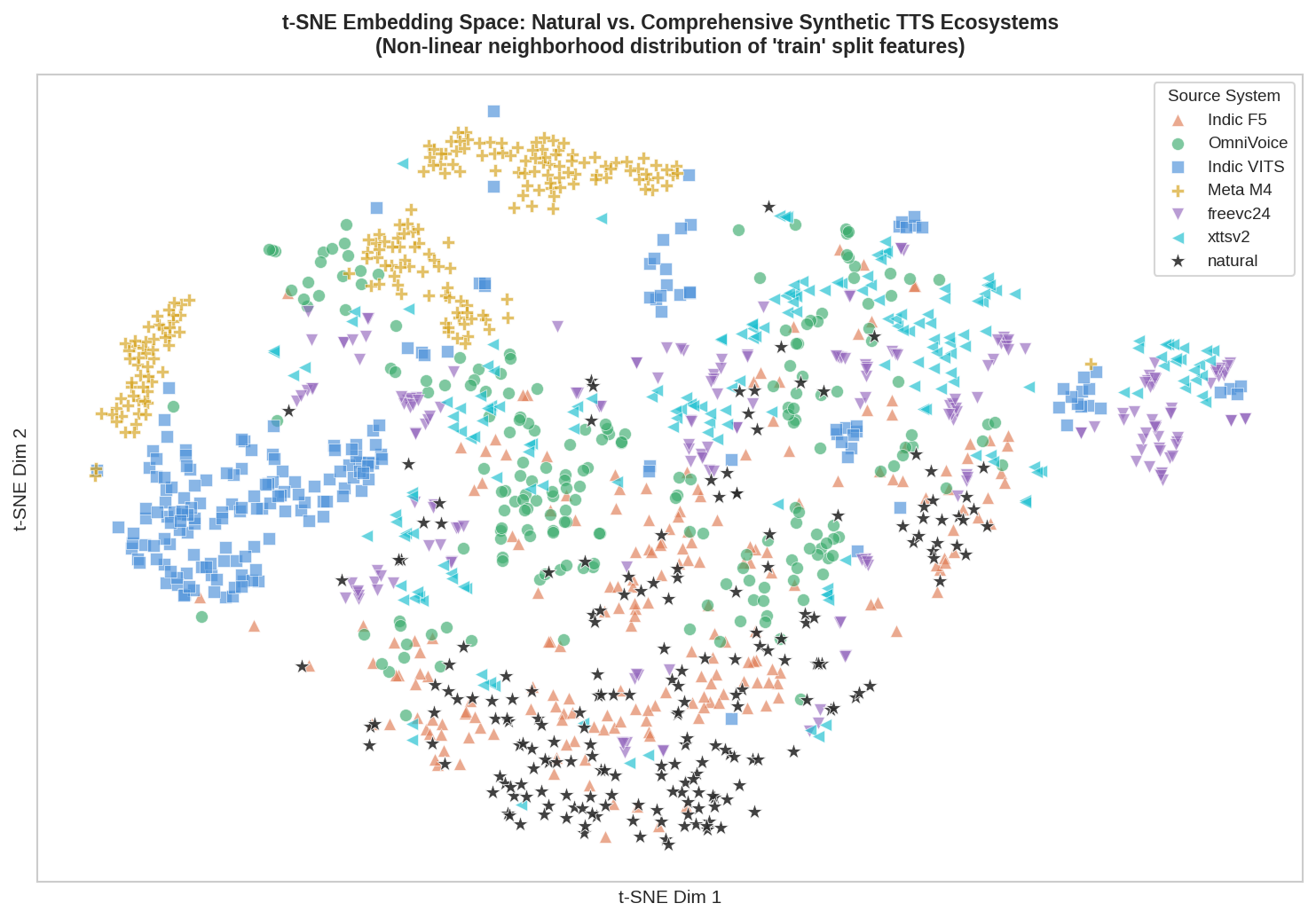}
    \caption{tsne plot}
    \label{fig:tsne_train_only}
\end{figure}

\begin{table}[ht]
\centering
\caption{Pairwise embedding similarity between Natural and Synthetic Speech Systems.}
\label{tab:pairwise_similarity}
\resizebox{\columnwidth}{!}{
\begin{tabular}{llcc}
\hline
\textbf{Model A} & \textbf{Model B} &
\textbf{Cosine Distance} &
\textbf{Euclidean Distance} \\
\hline

Natural & M4         & 0.0105 & 4.0726 \\
F5      & M4         & 0.0088 & 3.7342 \\
freevc24 & M4         & 0.0070 & 3.4127 \\
xttsv2  & M4         & 0.0055 & 2.9660 \\
Indic VITS & freevc24 & 0.0045 & 2.6941 \\
Indic VITS & xttsv2   & 0.0046 & 2.6779 \\
M4      & Omni       & 0.0044 & 2.6566 \\
Natural & Omni       & 0.0043 & 2.6043 \\
Natural & xttsv2     & 0.0040 & 2.5126 \\
Indic VITS & Omni    & 0.0040 & 2.5097 \\
Natural & Indic VITS & 0.0039 & 2.4883 \\
Indic VITS & M4      & 0.0038 & 2.4574 \\
Natural & freevc24   & 0.0035 & 2.4370 \\
F5      & Indic VITS & 0.0037 & 2.4170 \\
F5      & freevc24   & 0.0032 & 2.3300 \\
F5      & xttsv2     & 0.0029 & 2.1521 \\
F5      & Omni       & 0.0024 & 1.9494 \\
Omni    & freevc24   & 0.0022 & 1.9178 \\
Omni    & xttsv2     & 0.0010 & 1.2688 \\
freevc24 & xttsv2    & 0.0008 & 1.1953 \\
Natural & F5         & 0.0008 & 1.1217 \\

\hline
\end{tabular}
}
\end{table}

\section{Conclusion}

We present a unified study of spontaneous speech detection and synthetic speech generalisation across 22 Indic languages. Encoder choice matters: Wav2vec2 \citep{baevski2020wav2vec2} shows a language-discriminability/spontaneity trade-off absent in Whisper \citep{radford2022whisper} and Vaani \citep{pulikodan2026vaani}. Training diversity drives OOD deepfake generalisation, raising recall from 7\% to 51\% across four TTS systems, governed by embedding proximity to unseen synthetic speech rather than natural speech. Future work should extend TTS and language coverage, test fine-tuned encoders, probe spoof detectors for this same language trade-off, and explore joint spontaneity-authenticity modelling for robust real-world Indic detectors.

\bibliography{custom}            

@inproceedings{batliner1995,
  author    = {Batliner, Anton and Kompe, Ralf and Kie{\ss}ling, Andreas 
               and N{\"o}th, Elmar and Niemann, Heinrich},
  title     = {Can You Tell Apart Spontaneous and Read Speech if You Just Look at Prosody?},
  booktitle = {Speech Recognition and Coding},
  series    = {NATO ASI Series},
  volume    = {147},
  pages     = {321--324},
  publisher = {Springer},
  address   = {Berlin},
  year      = {1995},
  doi       = {10.1007/978-3-642-57745-1_47},
  url       = {https://link.springer.com/chapter/10.1007/978-3-642-57745-1_47}
}

@article{busso2008,
  author    = {Busso, Carlos and Bulut, Murtaza and Lee, Chi-Chun and 
               Kazemzadeh, Abe and Mower, Emily and Kim, Samuel and 
               Chang, Jeannette N. and Lee, Sungbok and Narayanan, Shrikanth S.},
  title     = {{IEMOCAP}: Interactive Emotional Dyadic Motion Capture Database},
  journal   = {Language Resources and Evaluation},
  volume    = {42},
  number    = {4},
  pages     = {335--359},
  year      = {2008},
  doi       = {10.1007/s10579-008-9076-6},
  url       = {https://link.springer.com/article/10.1007/s10579-008-9076-6}
}

@inproceedings{rouas2003,
  author    = {Rouas, Jean-Luc and Farinas, Jerome and Pellegrino, Fran{\c{c}}ois 
               and Andr{\'e}-Obrecht, R{\'e}gine},
  title     = {Modeling Prosody for Language Identification on Read and 
               Spontaneous Speech},
  booktitle = {Proceedings of the IEEE International Conference on Acoustics, 
               Speech, and Signal Processing (ICASSP)},
  volume    = {6},
  pages     = {40--43},
  publisher = {IEEE},
  year      = {2003},
  doi       = {10.1109/ICASSP.2003.1200912}
}

@article{nakamura2008,
  author    = {Nakamura, Masanobu and Iwano, Koji and Furui, Sadaoki},
  title     = {Differences Between Acoustic Characteristics of Spontaneous 
               and Read Speech and Their Effects on Speech Recognition 
               Performance},
  journal   = {Computer Speech \& Language},
  volume    = {22},
  number    = {2},
  pages     = {171--184},
  year      = {2008},
  doi       = {10.1016/j.csl.2007.07.003}
}

@inproceedings{christodoulides2020,
  author    = {Christodoulides, George},
  title     = {Speaking Style Prosodic Variation and the Prosody-Syntax 
               Interface: A Large-Scale Corpus Study},
  booktitle = {Proceedings of Speech Prosody 2020},
  pages     = {705--709},
  year      = {2020},
  doi       = {10.21437/SpeechProsody.2020-144}
}

@inproceedings{ryant2016,
  author    = {Ryant, Neville and Liberman, Mark},
  title     = {Automatic Analysis of Phonetic Speech Style Dimensions},
  booktitle = {Proceedings of Interspeech 2016},
  pages     = {77--81},
  publisher = {ISCA},
  year      = {2016},
  doi       = {10.21437/Interspeech.2016-1213}
}

@article{elisha2024,
  author    = {Elisha, Shahar and others},
  title     = {Classification of Spontaneous and Scripted Speech for Multilingual Audio},
  journal   = {arXiv preprint},
  year      = {2024},
  url       = {https://arxiv.org/abs/2412.11896}
}

@inproceedings{javed2024,
  author    = {Javed, Tahir and Nawale, Janki and George, Eldho and 
               Joshi, Sakshi and Bhogale, Kaushal and Kumar, Pratyush 
               and Khapra, Mitesh M.},
  title     = {{IndicVoices}: Towards Building an Inclusive Multilingual 
               Speech Dataset for {Indian} Languages},
  booktitle = {Findings of the Association for Computational Linguistics: ACL 2024},
  pages     = {10740--10782},
  publisher = {Association for Computational Linguistics},
  address   = {Bangkok, Thailand},
  year      = {2024},
  doi       = {10.18653/v1/2024.findings-acl.639},
  url       = {https://aclanthology.org/2024.findings-acl.639}
}

@inproceedings{jung2022,
  author    = {Jung, Jee-weon and Heo, Hee-Soo and Tak, Hemlata and 
               Shim, Hye-jin and Chung, Joon Son and Lee, Bong-Jin and 
               Yu, Ha-Jin and Evans, Nicholas},
  title     = {{AASIST}: Audio Anti-Spoofing Using Integrated 
               Spectro-Temporal Graph Attention Networks},
  booktitle = {Proceedings of ICASSP 2022},
  pages     = {6367--6371},
  publisher = {IEEE},
  year      = {2022},
  doi       = {10.1109/ICASSP43922.2022.9747766},
  url       = {https://arxiv.org/abs/2110.01200}
}

@inproceedings{tak2021,
  author    = {Tak, Hemlata and Patino, Jose and Todisco, Massimiliano 
               and Nautsch, Andreas and Evans, Nicholas and Larcher, Anthony},
  title     = {End-to-End Anti-Spoofing with {RawNet2}},
  booktitle = {Proceedings of ICASSP 2021},
  pages     = {6369--6373},
  publisher = {IEEE},
  address   = {Toronto, ON, Canada},
  year      = {2021},
  doi       = {10.1109/ICASSP39728.2021.9414234},
  url       = {https://arxiv.org/abs/2011.01108}
}

@article{li2024,
  author    = {Li, Yuankun and others},
  title     = {Where Are We in Audio Deepfake Detection? {A} Systematic 
               Analysis over Generative and Detection Models},
  journal   = {arXiv preprint},
  year      = {2024},
  url       = {https://arxiv.org/abs/2410.04324}
}

@inproceedings{sharma2025,
  author    = {Sharma, Divya V and Ekbote, Vijval and Gupta, Anubha},
  title     = {{IndicSynth}: A Large-Scale Multilingual Synthetic Speech 
               Dataset for Low-Resource {Indian} Languages},
  booktitle = {Proceedings of the 63rd Annual Meeting of the Association 
               for Computational Linguistics (Volume 1: Long Papers)},
  pages     = {22037--22060},
  publisher = {Association for Computational Linguistics},
  address   = {Vienna, Austria},
  year      = {2025},
  url       = {https://aclanthology.org/2025.acl-long.1070/}
}

@inproceedings{mangalam2018,
  author    = {Mangalam, Karttikeya and Guha, Tanaya},
  title     = {Learning Spontaneity to Improve Emotion Recognition in Speech},
  booktitle = {Proceedings of Interspeech 2018},
  publisher = {ISCA},
  year      = {2018},
  doi       = {10.21437/Interspeech.2018-1872},
  url       = {https://www.isca-archive.org/interspeech_2018/mangalam18_interspeech.html}
}

@techreport{radford2022whisper,
  title       = {Robust Speech Recognition via Large-Scale Weak Supervision},
  author      = {Radford, Alec and Kim, Jong Wook and Xu, Tao and Brockman, Greg and McLeavey, Christine and Sutskever, Ilya},
  institution = {OpenAI},
  year        = {2022},
  url         = {https://cdn.openai.com/papers/whisper.pdf}
}

@inproceedings{baevski2020wav2vec2,
  title     = {wav2vec 2.0: A Framework for Self-Supervised Learning of Speech Representations},
  author    = {Baevski, Alexei and Zhou, Yuhao and Mohamed, Abdelrahman and Auli, Michael},
  booktitle = {Advances in Neural Information Processing Systems (NeurIPS)},
  volume    = {33},
  pages     = {12449--12460},
  year      = {2020},
  url       = {https://arxiv.org/abs/2006.11477}
}

@inproceedings{gong2021ast,
  title     = {AST: Audio Spectrogram Transformer},
  author    = {Gong, Yuan and Chung, Yu-An and Glass, James},
  booktitle = {Proceedings of Interspeech 2021},
  pages     = {571--575},
  year      = {2021},
  doi       = {10.21437/Interspeech.2021-698},
  url       = {https://arxiv.org/abs/2104.01778}
}

@inproceedings{chen2023beats,
  title     = {BEATs: Audio Pre-Training with Acoustic Tokenizers},
  author    = {Chen, Sanyuan and Wu, Yu and Wang, Chengyi and Liu, Shujie and Tompkins, Daniel and Chen, Zhuo and Che, Wanxiang and Yu, Xiangzhan and Wei, Furu},
  booktitle = {Proceedings of the 40th International Conference on Machine Learning (ICML)},
  pages     = {5178--5193},
  year      = {2023},
  publisher = {PMLR},
  url       = {https://proceedings.mlr.press/v202/chen23ag.html}
}

@article{pulikodan2026vaani,
  title   = {VAANI: Capturing the Language Landscape for an Inclusive Digital India},
  author  = {Pulikodan, Sujith and Singh, Abhayjeet and Basu, Agneedh and Desai, Nihar and Kumar J, Pavan and Bhat, Pranav D and Dharmaraju, Raghu and Gupta, Ritika and Udupa, Sathvik and Kumar, Saurabh and Sharma, Sumit and Sanka, Visruth and Tewari, Dinesh and Dhand, Harsh and Kamat, Amrita and Singh, Sukhwinder and Vashishth, Shikhar and Talukdar, Partha and Acharya, Raj and Ghosh, Prasanta Kumar},
  journal = {arXiv preprint arXiv:2603.28714},
  year    = {2026},
  url     = {https://arxiv.org/abs/2603.28714}
}

@inproceedings{todisco2017cqcc,
  author    = {Todisco, Massimiliano and Delgado, Héctor and Evans, Nicholas},
  title     = {Constant Q Cepstral Coefficients: A Spoofing Countermeasure for Automatic Speaker Verification},
  booktitle = {Proceedings of Interspeech 2017},
  pages     = {23--27},
  year      = {2017},
  doi       = {10.21437/Interspeech.2017-1111},
  url       = {https://www.isca-speech.org/archive/Interspeech_2017/pdfs/1111.PDF}
}

@inproceedings{wu2015lfcc,
  author    = {Wu, Zhizheng and Kinnunen, Tomi and Evans, Nicholas and Yamagishi, Junichi and Hanilçi, Cemal and Sahidullah, Md and Alegre, Federico},
  title     = {Spoofing and Countermeasures for Speaker Verification: A Survey},
  booktitle = {Speech Communication},
  volume    = {66},
  pages     = {130--153},
  year      = {2015},
  doi       = {10.1016/j.specom.2014.10.005},
  url       = {https://www.sciencedirect.com/science/article/pii/S0167639314001080}
}

@inproceedings{chen2022wavlm,
  author    = {Chen, Sanyuan and Wang, Chengyi and Chen, Zhuo and Wu, Yu and Jia, Shujie and Hua, Wei and Wang, Dong and Zhou, Ming and Li, Jinyu},
  title     = {WavLM: Large-Scale Pre-Training for Full Stack Speech Processing},
  booktitle = {Proceedings of NeurIPS 2022},
  year      = {2022},
  url       = {https://arxiv.org/abs/2110.13900}
}

@misc{vitsrasa2024,
  author       = {{AI4Bharat Team}},
  title        = {{VITS} {TTS} for {Indian} Languages ({vits\_rasa\_13})},
  year         = {2024},
  howpublished = {\url{https://huggingface.co/ai4bharat/vits_rasa_13}},
  publisher    = {Hugging Face}
}

@misc{indicf52025,
  author       = {S~V, Praveen and Anand, Srija and Siddhartha, Soma and Khapra, Mitesh M.},
  title        = {{IndicF5}: High-Quality Text-to-Speech for {Indian} Languages},
  year         = {2025},
  howpublished = {\url{https://github.com/AI4Bharat/IndicF5}},
  publisher    = {AI4Bharat},
  note         = {GitHub / Hugging Face model release}
}

@inproceedings{li2023freevc,
  author    = {Li, Jingyi and Tu, Weiping and Xiao, Li},
  title     = {{FreeVC}: Towards High-Quality Text-Free One-Shot Voice Conversion},
  booktitle = {Proceedings of ICASSP 2023},
  publisher = {IEEE},
  year      = {2023},
  url       = {https://arxiv.org/abs/2210.15418}
}

@inproceedings{casanova2024xtts,
  author    = {Casanova, Edresson and Davis, Kelly and G{\"o}lge, Eren 
               and G{\"o}knar, G{\"o}rkem and Gulea, Iulian and Hart, Logan 
               and Aljafari, Aya and Meyer, Joshua and Morais, Reuben 
               and Olayemi, Samuel and Weber, Julian},
  title     = {{XTTS}: A Massively Multilingual Zero-Shot Text-to-Speech Model},
  booktitle = {Proceedings of Interspeech 2024},
  pages     = {4978--4982},
  year      = {2024},
  doi       = {10.21437/Interspeech.2024-2016}
}

@article{zhu2026omnivoice,
  author  = {Zhu, Han and Ye, Lingxuan and Kang, Wei and Yao, Zengwei 
             and Guo, Liyong and Kuang, Fangjun and Han, Zhifeng 
             and Zhuang, Weiji and Lin, Long and Povey, Daniel},
  title   = {{OmniVoice}: Towards Omnilingual Zero-Shot Text-to-Speech with 
             Diffusion Language Models},
  journal = {arXiv preprint arXiv:2604.00688},
  year    = {2026},
  url     = {https://arxiv.org/abs/2604.00688}
}

@inproceedings{asvspoof2019,
  author    = {Wang, Xin and Kinnunen, Tomi and Sahidullah, Md and Delgado, Héctor and Evans, Nicholas and Yamagishi, Junichi and Todisco, Massimiliano},
  title     = {ASVspoof 2019: A Large-Scale Database for Spoofing Attack Countermeasures},
  booktitle = {Proceedings of Interspeech 2019},
  pages     = {100--104},
  year      = {2019},
  doi       = {10.21437/Interspeech.2019-2249},
  url       = {https://www.isca-speech.org/archive/Interspeech_2019/pdfs/2249.pdf}
}

@inproceedings{asvspoof2021,
  author    = {Yamagishi, Junichi and Wang, Xin and Todisco, Massimiliano and Patino, Jose and Nautsch, Andreas and Evans, Nicholas and Kinnunen, Tomi and Sahidullah, Md and Delgado, Héctor},
  title     = {ASVspoof 2021: Automatic Speaker Verification Spoofing and Countermeasures Challenge Evaluation Plan},
  booktitle = {Proceedings of Interspeech 2021},
  pages     = {4203--4207},
  year      = {2021},
  doi       = {10.21437/Interspeech.2021-0003},
  url       = {https://www.isca-speech.org/archive/pdfs/interspeech_2021/0003.pdf}
}

@inproceedings{asvspoof2023,
  author    = {Yamagishi, Junichi and Todisco, Massimiliano and Wang, Xin and Nautsch, Andreas and Evans, Nicholas and Patino, Jose and Kinnunen, Tomi and Sahidullah, Md and Delgado, Héctor},
  title     = {ASVspoof 2023: The Automatic Speaker Verification Spoofing and Countermeasures Challenge},
  booktitle = {Proceedings of Interspeech 2023},
  pages     = {1234--1238},
  year      = {2023},
  doi       = {10.21437/Interspeech.2023-0003},
  url       = {https://www.isca-speech.org/archive/interspeech_2023/0003.pdf}
}

@article{barrault2023seamless,
  author  = {Barrault, Lo{\"i}c and Chung, Yu-An and Coria Meglioli, Mariano 
             and Dale, David and Dong, Ning and Duppenthaler, Mark 
             and Duquenne, Paul-Ambroise and Ellis, Brian and Elsahar, Hady 
             and Haaheim, Justin and others},
  title   = {Seamless: Multilingual Expressive and Streaming Speech Translation},
  journal = {arXiv preprint arXiv:2312.05187},
  year    = {2023},
  url     = {https://arxiv.org/abs/2312.05187}
}

\end{document}